\documentclass[notitlepage,nofootinbib,amssymb, nobibnotes, aps, prd,showpacs,showkeys]{revtex4-1}
\usepackage[english]{babel}
\usepackage{amsfonts}
\usepackage{amsthm}
\usepackage{graphicx}
\usepackage{braket}
\usepackage{amssymb}
\usepackage{hyperref}
\usepackage{xspace}
\usepackage{gensymb}
\usepackage{amsmath}
\usepackage[usenames,dvipsnames,svgnames,table]{xcolor}
\usepackage{bm}
\usepackage[normalem]{ulem}

\renewcommand{\epsilon}{\varepsilon}

\newcommand{\chicomega}{\ensuremath{\chi_{c0}\omega}\xspace}

\newcommand{\hcpipi}{\ensuremath{h_c\pi^+\pi^-}\xspace}
\newcommand{\Prob}{\ensuremath{\text{Prob}}\xspace}
\newcommand{\ie}{{\it i.e.}\xspace}
\newcommand{\BR}{\ensuremath{\mathcal{B}}\xspace}

\begin{document}

\title{A note on the newly observed $Y(4220)$ resonance}
\author{R.~Faccini$^{*,\P}$, G.~Filaci$^{*}$, A.L.~Guerrieri$^{\dag}$, A.~Pilloni$^{*,\P}$, A.D.~Polosa$^{*,\P}$}
\affiliation{
\mbox{$^*$Dipartimento di Fisica, ``Sapienza'' Universit\`a di Roma, P.le A. Moro 2, I-00185 Roma, Italy}\\
\mbox{$^\P$INFN sez. Roma 1, P.le A. Moro 2, I-00185 Roma, Italy}\\
\mbox{$^\dag$Dipartimento di Fisica and INFN, Universit\`a di Roma ``Tor Vergata'',}\\ \mbox{Via della Ricerca Scientifica 1, I-00133 Roma, Italy}
 }

\begin{abstract}BES~III Collaboration has recently observed a vector resonance in the \chicomega channel, at a mass of about $4220$~MeV, named $Y(4220)$. Hints of a similar structure appear in the \hcpipi channel. We find that the two observations are likely due to the same state, which we identify with one of the expected diquark-antidiquark resonances with orbital quantum number $L=1$. This assignment fulfills heavy quark spin conservation. The measured branching ratio of the $Y(4220)$ into \chicomega and \hcpipi is compatible with the prediction for such a tetraquark state.
\end{abstract}

\pacs{14.40.Rt, 12.39.Jh, 13.25.Gv}

\maketitle

In a very recent paper, the BES~III Collaboration reports the $e^+e^- \to \chi_{cJ} \omega$ ($J=0,1,2$) production cross section as a function of $\sqrt{s}$~\cite{beschicomega}. 
Hints of a resonant structure  are present in the \chicomega channel at $\sim 30$~MeV above threshold ({\it i.e.} at about $4220$~MeV), whereas no evident structure appears in the $\chi_{c1,2} \omega$ channels. Some theoretical interpretations for this peak have been proposed~\cite{altri}.
BES Collaboration  also reported the measurement of $e^+e^- \to \hcpipi$ production cross section as a function of $\sqrt{s}$~\cite{beshcpipi}. Hints of structures not compatible with the $Y(4260)$ have been found~\cite{hadrocharmonium,yuan}: in particular a narrow peak at $\sim 4220$~MeV.
 Heavy quark spin symmetry prevents any ordinary charmonium from decaying into both $\chi_c$ and $h_c$. Violations of this symmetry have already been observed in the bottomonium sector, and are explained in~\cite{HQSvoloshin,alivecchio,alimaiani}.
In the charmonium mass region, many exotic charmonium-like states have been identified according to the diquark-antidiquark model~\cite{tetraquarks} (for a review, see~\cite{review}).
In particular, the latest model~\cite{Maiani:2014aja} predicts a tetraquark state, named $Y_3$, with quantum numbers $J^{PC}=1^{--}$, and mass and decay modes compatible with a $Y(4220)$ resonance. The wave function of this tetraquark state contains both heavy quark spin states, so it can naturally decay into both $\chi_{c0} \omega$ and $\hcpipi$ with no violation of the heavy quark spin. 
Since the Breit-Wigner parameters of the peaks measured in the two channels \chicomega and \hcpipi are very similar, we test the hypothesis that the two observed structures may coincide.

We fit data with two different models (I and II in the following) similar to those considered in Refs.~\cite{yuan,beschicomega}. In the \hcpipi invariant mass distribution, we add to the BES dataset the experimental point $\sigma_{\hcpipi}\left(4.17\text{~GeV}\right)=(15.6 \pm 4.2)$~pb\footnote{$\sigma_f(m)$ indicates the cross section $\sigma(e^+e^- \to f)$ at $\sqrt{s}=m$.} by \mbox{CLEO-$c$~\cite{cleohcpipi}}, with statistical and systematic errors added in quadrature.
For the BES data, we take into account only statistical errors, since the systematic ones are common to all points and are not expected to modify the shape of the distribution. 

Following model-I, we fit the \hcpipi and \chicomega data with the sum of a Breit-Wigner corrected for the energy dependence given by PCAC, and  a pure phase-space background. 
To test our hypothesis, the mass and the width of the resonance are constrained to be the same in both channels. Thus, the fitting functions are:
\begin{align}
 \sigma_{\hcpipi}(m) &= \left|A \sqrt{\text{PS}_3(m)} + B e^{i\phi_1} \sqrt{\frac{\text{PS}^\prime_3(m)}{\text{PS}^\prime_3(m_0)}}\,\text{BW}\!\left(m,m_0,\Gamma\right)  \right|^2 ,\\
 \sigma_{\chicomega}(m) &= \left|C + \frac{D e^{i\phi_2}}{\sqrt{\text{PS}_2(m_0)}}\,\text{BW}\!\left(m,m_0,\Gamma\right) \right|^2 \text{PS}_2(m),
\end{align}
where $m_0$ and $\Gamma$ are the mass and width of the resonance, $m$ is the invariant mass of the system, $\text{BW}\!\left(m,m_0,\Gamma\right)= \left(m^2 - m_0^2 + i m_0 \Gamma\right)^{-1}$, $B=\sqrt{12\pi \mathcal{B}_{\hcpipi} \Gamma_{ee} \Gamma}$, $D=\sqrt{12\pi \mathcal{B}_{\chicomega} \Gamma_{ee} \Gamma}$, $\text{PS}_n$ is the $n$-body phase space, and $\text{PS}^\prime_3$ is the PCAC-corrected phase space~\cite{HQSvoloshin,soviet}, namely:
\begin{equation}
 \text{PS}^\prime_3 \propto \int d\Phi_3 \left(E^+ p^- + E^- p^+\right)^2,
\end{equation}
where $E^\pm$ ($p^\pm$) is the energy (momentum) of $\pi^\pm$ in the CM frame.
With this model, we get a mass of $4213 \pm 12$~MeV and a width of $52 \pm 24$~MeV. The $\chi^2 / \text{DOF} = 17.38 / 15 $, corresponding to a $\Prob(\chi^2) = 30\%$ (see \figurename{~\ref{fig:oneresonance} and \tablename{~\ref{tab:oneresonance}}). The fit gives two distinct solutions for the Breit-Wigner amplitudes, corresponding to a constructive and destructive interference in the \hcpipi channel, respectively.\footnote{This ambiguity does not affect the \chicomega channel, being the background compatible with zero.} 

To obtain the significance of the $Y(4220)$, we perform a likelihood ratio test: we repeat the fit according to a pure phase-space background hypothesis, {\it i.e.} forcing $B=D=0$. The $\Delta \chi^2 / \Delta\text{DOF}$ with respect to the full fit is $131 / 6$, which rejects the pure background hypothesis with a significance $> 10\sigma$.

By comparing the Breit-Wigner amplitudes in the two channels, we get the ratio:
\begin{subequations}
\begin{align}
 \frac{\BR\left(Y(4220)\to\chicomega \right)}{\BR\left(Y(4220)\to\hcpipi \right)} &= 8.3\pm 4.8 \pm 1.9\quad\text{(Sol. A)}\label{ratexpbuono}\\
 &= 0.48 \pm 0.20  \pm 0.11\quad\text{(Sol. B)}
\end{align}\label{ratexp}
\end{subequations}
where the second error is the quadrature sum of the systematic uncertainties of $15\%$ for $\sigma_{\chicomega}$~\cite{beschicomega} and $18\%$ for $\sigma_{\hcpipi}$~\cite{beshcpipi}. In this way, we consider the two BES datasets to have statistically independent systematics, which leads to a conservative estimate of the error.~\footnote{We remark that the ratio in Eq.~\eqref{ratexpbuono} is compatible with the ratio of the branching fractions of Ref.~\cite{beschicomega} and \cite{yuan}, once the value for $\BR(Y\to \hcpipi) \times \Gamma_{ee}$ in Ref.~\cite{yuan} is corrected by a typo of one order of magnitude~\cite{PC}.} 

\begin{figure}[t]
\centering
\includegraphics[width=0.45\textwidth]{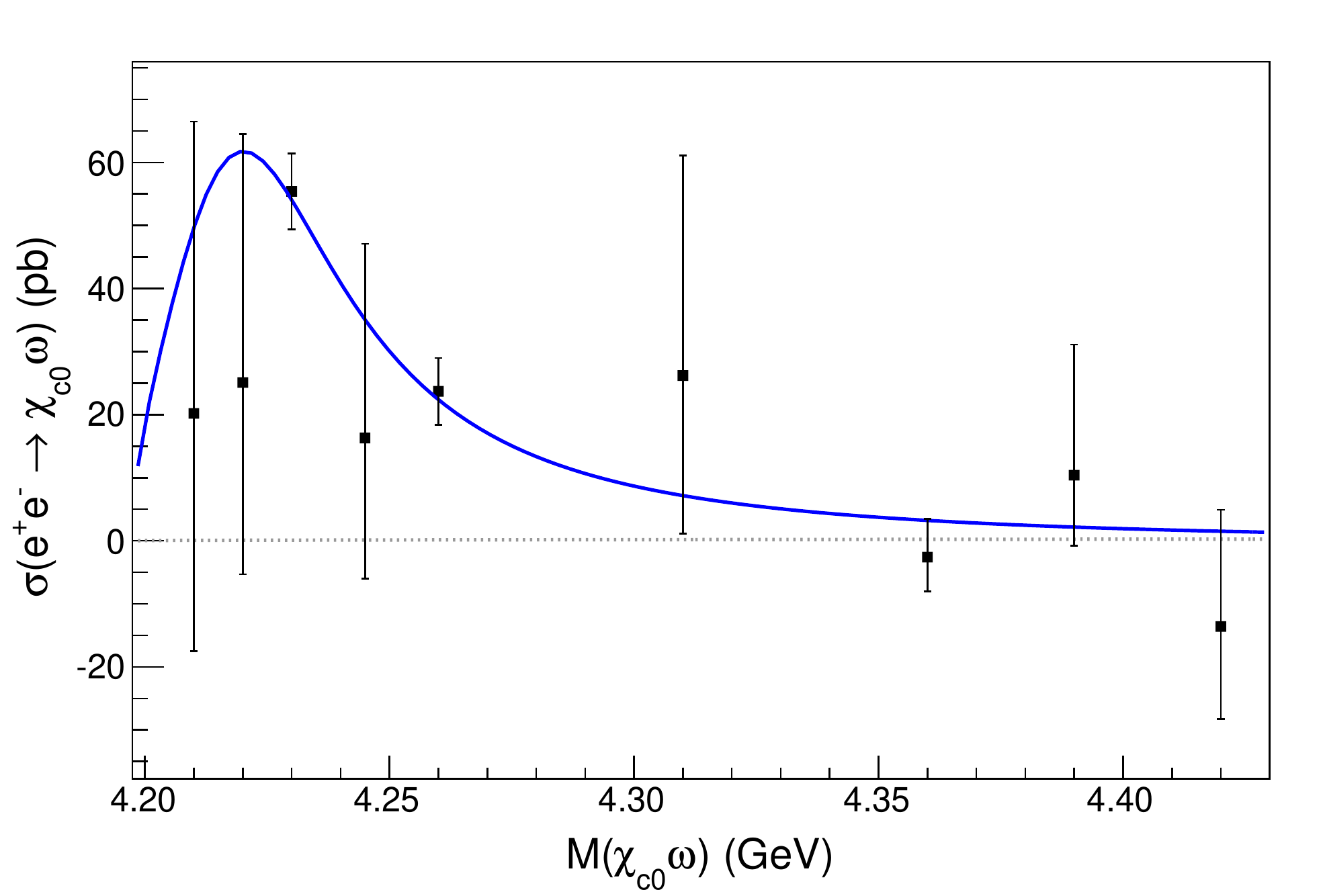}
\includegraphics[width=0.45\textwidth]{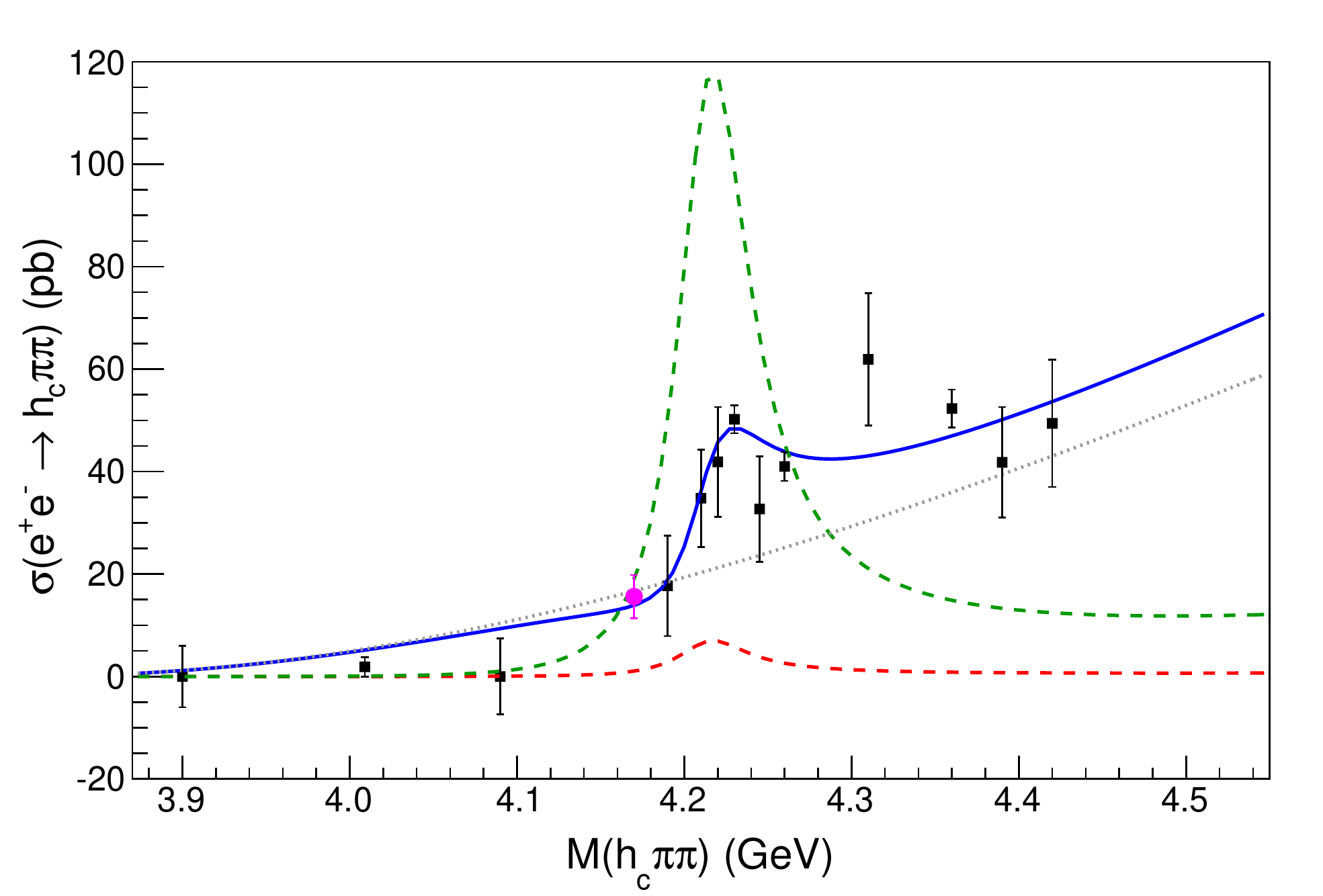}
\caption{Combined fits of \chicomega~\cite{beschicomega} and $h_c\pi^+\pi^-$ data~\cite{beshcpipi,cleohcpipi}: model-I. The purple disk in the right panel is the CLEO-$c$ data point at $\sqrt{s}=4.17$~GeV. The red (smaller) and the green (larger) dashed curves are the Breit-Wigner curves for solution A and B, respectively. The dotted gray curve is background.}
 \label{fig:oneresonance}
\end{figure}

\begin{table}[t]
\begin{tabular}{c|c c}
\hline\hline
& Solution A & Solution B\\ \hline
$A$ & \multicolumn{2}{c}{$(3.50 \pm 0.36) \!\times\! 10^{-2} \,\text{GeV}^{-2}$ }\\
$C$ & \multicolumn{2}{c}{$(0.5 \pm 2.3)\! \times\! 10^{-4} \,\text{GeV}^{-1}$}\\
$\mathcal{B}_{\hcpipi} \times \mathcal{B}_{ee}$ & \parbox[c]{4cm}{$(8.3 \pm 3.4) \times 10^{-9}$} & \parbox[c]{4cm}{$(1.41 \pm 0.27)\times 10^{-7}$}\\
$\mathcal{B}_{\chicomega} / \mathcal{B}_{\hcpipi}$ & $8.3 \pm 4.8$ & $0.48 \pm 0.20$ \\
$m_0$ & \multicolumn{2}{c}{$(4213 \pm 12)\, \text{MeV}$} \\
$\Gamma$ & \multicolumn{2}{c}{$(52 \pm 24)\, \text{MeV}$} \\
$\phi_1$ & $(32 \pm 19)\degree$ & $(276.1 \pm 5.4)\degree$ \\
$\phi_2$ & \multicolumn{2}{c}{$(182 \pm 240)\degree$} \\
\hline
$\chi^2/DOF$ & \multicolumn{2}{c}{$17.38/15$} \\ \hline\hline
\end{tabular}
\caption{Results of the fit (model-I). The phase-space background in \chicomega distribution is compatible with zero, hence the large error on $\phi_2$.}
 \label{tab:oneresonance}
\end{table}

According to model-II, the background is parametrized by a broad Breit-Wigner:
\begin{align}
 \sigma_{\hcpipi}(m) &= \left|\frac{B_1}{\sqrt{\text{PS}^\prime_3(m_1)}}\text{BW}\left(m,m_1,\Gamma_1\right)+\frac{B_2 e^{i\phi_1}}{\sqrt{\text{PS}^\prime_3(m_2)}}\text{BW}\left(m,m_2,\Gamma_2\right)\right|^2 \text{PS}_3(m),\\
 \sigma_{\chicomega}(m) &= \left|C + \frac{D e^{i\phi_2}}{\sqrt{\text{PS}_2(m_1)}}\text{BW}\left(m,m_1,\Gamma_1\right) \right|^2 \text{PS}_2(m).
\end{align}
With this model, we get a mass of $4234 \pm 6$~MeV and a width of $34 \pm 16$~MeV. The $\chi^2 / \text{DOF} = 6.26 / 13 $, corresponding to a $\Prob(\chi^2) =93.6\%$. 
In this case, the significance of the signal is $>9 \sigma$. This model yields to a branching fraction ratio of 
$\BR\left(Y\to\chicomega \right)/\BR\left(Y\to\hcpipi \right) = 6.0 \pm 8.9 \pm 1.4\, \text{(Sol. A)} = 0.32 \pm 0.23 \pm  0.07\,\text{(Sol. B)}$ -- see \figurename{~\ref{fig:tworesonaces}} and \tablename{~\ref{tab:tworesonaces}}.

Even though model-II fits data better, the presence of two peaks with so different widths and amplitudes appears unlikely.
The broad Breit-Wigner peak just acts as a more effective (but less plausible) parameterization of the background although more data at masses higher than $4.5$~GeV are needed to discriminate experimentally between the two models. The fitted values of mass and width of  the $Y(4220)$ according to two models are not in statistical agreement since the two data samples are the same. Nonetheless, we will assume that the best estimates of the $Y(4220)$ mass and width come  from the model-I fit since it is sounder from the physical point of view. In any case, the conclusions of our study would be the same considering the results of the fit to  model-II, albeit larger errors. 

\begin{figure}[t]
\centering
\includegraphics[width=0.45\textwidth]{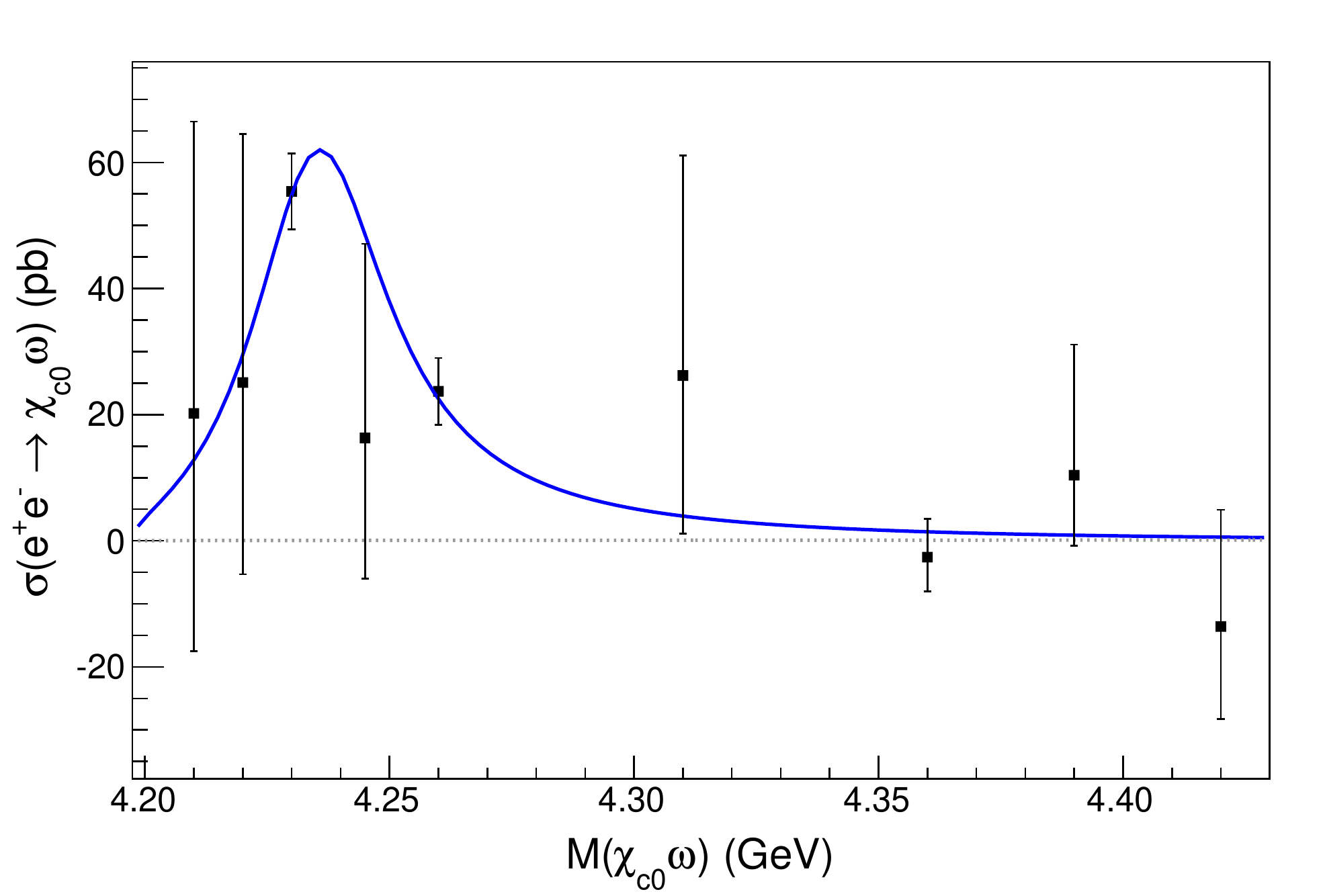}
\includegraphics[width=0.45\textwidth]{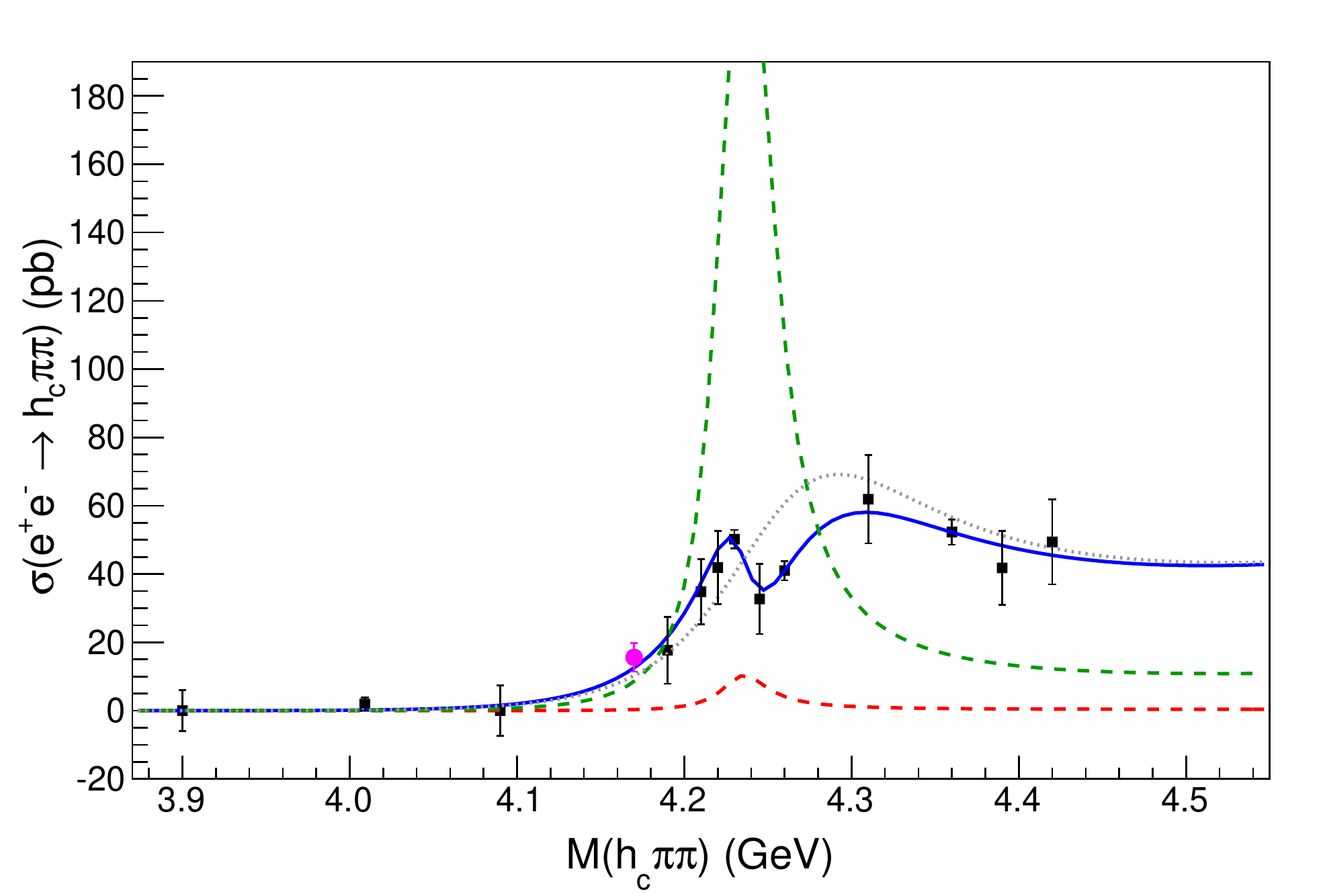}
\caption{Combined fits of \chicomega~\cite{beschicomega} and $h_c\pi^+\pi^-$ data~\cite{beshcpipi,cleohcpipi}: model-II. The purple disk in the right panel is the CLEO-$c$ data point at $\sqrt{s}=4.17$~GeV. The red (smaller) and the green (larger) dashed curves are the Breit-Wigner curves for solution A and B, respectively. The dotted gray curve is a phase-space background in  \chicomega, and a second broad Breit-Wigner in \hcpipi.}
\label{fig:tworesonaces}
\end{figure}

\begin{table}[t]
\begin{tabular}{c|c c}
\hline\hline
 & Solution A & Solution B \\ \hline
$C$ & \multicolumn{2}{c}{$(0.3 \pm 2.3) \times10^{-4} \,\text{GeV}^{-1}$} \\
$\left(\mathcal{B}_{\hcpipi} \times \mathcal{B}_{ee}\right)_1$ & \parbox[c]{4cm}{$(1.3 \pm 1.5) \times 10^{-8}$} & \parbox[c]{4cm}{$(3.3 \pm 1.2) \times 10^{-7}$} \\
$\left(\mathcal{B}_{\hcpipi} \times \mathcal{B}_{ee}\right)_2$ & $(7.2 \pm 3.6) \times 10^{-8}$ & $(1.6 \pm 1.3) \times 10^{-7}$ \\
$\left(\mathcal{B}_{\chicomega} / \mathcal{B}_{\hcpipi}\right)_1$ & $6.0 \pm 8.9$ & $0.23 \pm 0.18$ \\
$m_1$ & \multicolumn{2}{c}{$(4234.4 \pm 5.7)\, \text{MeV}$} \\
$\Gamma_1$ & \multicolumn{2}{c}{$(34 \pm 16)\, \text{MeV}$} \\
$m_2$ & \multicolumn{2}{c}{$(4255 \pm 18)\, \text{MeV}$} \\
$\Gamma_2$ & \multicolumn{2}{c}{$(158 \pm 52)\, \text{MeV}$} \\
$\phi_1$ & $(86 \pm 42)\degree$ & $(160 \pm 15)\degree$  \\
$\phi_2$ & \multicolumn{2}{c}{undetermined} \\
\hline
$\chi^2/DOF$ &  \multicolumn{2}{c}{$6.26/13$} \\ \hline\hline
\end{tabular}
\caption{Results of the fit (model-II). The phase-space background in \chicomega distribution is compatible with zero, hence the large error on $\phi_2$}
 \label{tab:tworesonaces}
\end{table}

Ref.~\cite{feshbach} proposed that exotic resonances are due to a Feshbach mechanism, \ie a resonance appears in an open (molecular) channel because of the hybridization with a closed channel (discrete level of a tetraquark Hamiltonian). 
The width of these resonances can be evaluated to be $\Gamma = A \sqrt{M - M_{th}}$, being $M_{th}$ the mass of the closest 2-body threshold, and $A = 10 \pm 5~\text{GeV}^{1/2}$. This formula predicts a width $\Gamma=48 \pm 32$~MeV for the $Y(4220)$, compatible with the experimental one. 

To further check the predictions within the tetraquark model~\cite{Maiani:2014aja}, 
we compute the ratio in Eq.~\eqref{ratexp}. 
The same analysis of the $\hcpipi$ final state showed a resonance, dubbed  $Z_c^\prime(4020)$, in the $e^+e^-\to Z_c^\prime(4020)^\pm \pi^\mp \to \hcpipi$ process~\cite{beshcpipi}.  
From the cross sections in Ref.~\cite{beshcpipi}, we can see that the fractions $R_Z=\sigma\left(e^+e^- \to Z_c^{\prime\pm} \pi^\mp \to \hcpipi\right) / \sigma\left(e^+e^- \to \hcpipi\right)$ at $\sqrt{s} = 4.23$, $4.26$ and $4.36$~GeV do not vary with $\sqrt{s}$ (see \tablename{~\ref{tab:Zfrac}}). The first point is very close to the $Y(4220)$ peak, and the other ones are slightly above. This would suggest that the same fraction occurs in the resonant events $R_{YZ} = \sigma\left(Y \to Z_c^{\prime\pm} \pi^\mp \to \hcpipi\right) / \sigma\left(Y \to \hcpipi\right)$. We therefore can preliminary assume $R_{YZ} = R_Z(\sqrt{s} = 4.23\text{ GeV}) = (17 \pm 7)\%$. However, we remark that we have no information on $R_Z$ in the left sideband, and a proper multidimensional analysis is due to better establish $R_{YZ}$. In the following, we will show our results as a function of $R_{YZ}$. 
On the other hand, we will not include an intermediate $Z_c(3900)^+ \pi^-$ channel, since the signal $Z_c(3900)^+\to h_c \pi^+$ is not significant. We also estimate the contribution of a $\pi\pi$ resonance, in particular $Y \to h_c \sigma \to h_c \pi^+ \pi^-$, whose presence will be verified by a detailed Dalitz analysis when new data will by available by BES~III.

\begin{table}[b]
 \centering
 \begin{tabular}{c|c|c|c}
  $\sqrt{s}$ (GeV) & $\sigma\left(e^+e^- \to Z_c^{\prime\pm} \pi^\mp \to \hcpipi\right)$ (pb) & $\sigma\left(e^+e^- \to \hcpipi\right)$ (pb) & $R_Z$ (\%) \\ \hline
  $4.23$ & $8.7 \pm 1.9 \pm 2.8 \pm 1.4$ & $50.2 \pm 2.7 \pm 4.6 \pm 7.9$ & $17 \pm 7$\\
  $4.26$ & $7.4 \pm 1.7 \pm 2.1 \pm 1.2$ & $41.0 \pm 2.8 \pm 3.7 \pm 6.4$ & $18\pm 7$ \\
  $4.36$ & $10.3 \pm 2.3 \pm 3.1 \pm 1.6$ & $52.3 \pm 3.7 \pm 4.8 \pm 8.2$ & $20 \pm 8$\\
 \end{tabular}
\caption{Cross sections measured at BES~III at different $\sqrt{s}$~\cite{beshcpipi}. For $R_Z$, the systematics which do not cancel in the ratio are taken into account and summed in quadrature with the statistical error. }
\label{tab:Zfrac}
\end{table}

We parametrize the matrix elements by enforcing Lorentz invariance and discrete symmetries,
\begin{subequations}\begin{align}
 \left\langle \chi_{c0}(p)\,\omega(\eta,q) | Y(\lambda,P)\right\rangle &= g_\chi \,\eta\cdot \lambda,\\
 \left\langle Z_c^\prime(\eta,q)\,\pi(p) | Y(\lambda,P)\right\rangle &= g_Z \, \eta \cdot \lambda \frac{P\cdot p}{f_\pi M_Y},\label{eq:PCAC}\\
 \left\langle h_c(\eta,q)\,\sigma(p) | Y(\lambda,P)\right\rangle &= g_h \,\epsilon_{\mu\nu\rho\sigma} \eta^\mu \lambda^\nu \frac{P^\rho q^\sigma}{P\cdot q},\\
\left\langle \pi(q)\pi(p) | \sigma(P)\right\rangle &= \frac{P^2}{2f_\pi},\label{eq:sigpipi}
\end{align}\end{subequations}%
where $g_Z$, $g_h$ and $g_\chi$ are effective strong couplings with dimension of a mass.
Applying the reduction formula to the (off-shell) interpolating field of the pion, one obtains
\begin{align}
\left\langle \beta\,\pi\,|\,\alpha\right\rangle \to -\frac{1}{f_\pi} \left\langle \beta|\partial\cdot A(0)|\alpha\right\rangle \to -\frac{p^\mu_\pi}{f_\pi} \left\langle \beta|A_\mu(0)|\alpha\right\rangle
\label{eq:axial}
\end{align}
in the chiral limit. In our case, the latter matrix element is a vector, being $\alpha$ a vector and $\beta$ an axial-vector. Thus it is either a polarization or a momentum of $\alpha$, $\beta$. An $S$-wave transition is obtained in the latter case, Eq.~\eqref{eq:PCAC}. Similarly, the emission of two pions implies a factor $P^2$ in the amplitude $\sigma\to\pi\pi$, Eq.~\eqref{eq:sigpipi}.

Hence, the decay widths in narrow width approximation~\cite{strong} are:
\begin{subequations}
\begin{align}
\Gamma\left(Y(4220)\to\chicomega\right) &= \frac{1}{3} \frac{p^*(M_Y,m_{\chi},m_\omega)}{8\pi M_Y^2} g_{\chi}^2 \left(3 + \frac{p^{*2}(M_Y,m_\chi,m_\omega)}{m_\omega^2}\right),\\
\Gamma\left(Y(4220)\to Z_c^{\prime\pm}\pi^{\mp}\to\hcpipi\right) &= 2\times\frac{1}{3} \frac{g_Z^2}{8\pi M_Y^2} \int_{\left(m_{\pi} + m_h\right)^2}^{\left(M_Y - m_{\pi}\right)^2} \,ds \,p^*(M_Y,\sqrt{s},m_{\pi})\left ( 3+\frac{p^{*2}(M_Y,\sqrt{s},m_{\pi})}{s} \right) \nonumber\\
& \qquad\times
\frac{E^2_\pi(\sqrt{s})}{f^2_\pi}\,\frac{1}{\pi}\frac{m_Z \Gamma_Z}{(s-m_Z^2)^2 + m_Z^2 \Gamma_Z^2}\,
\frac{p^{*3}(\sqrt{s},m_h,m_{\pi})}{p^{*3}(m_Z,m_h,m_{\pi})}\,\frac{m_Z^3}{s^{3/2}}\,\BR\left(Z^\prime_c \to h_c \,\pi\right), \label{widthZ}\\
\Gamma\left(Y(4220)\to h_c\sigma\to\hcpipi\right) &= \frac{1}{3} \frac{g_h^2}{8\pi M_Y^2} \int_{4m_{\pi}^2}^{\left(M_Y - m_h\right)^2} \,ds\,p^{*}(M_Y,\sqrt{s},m_h)\,\frac{2p^{*2}(M_Y,\sqrt{s},m_h)}{m_h^2 + p^{*2}(M_Y,\sqrt{s},m_h)}\nonumber\\
 &\qquad\times
 \frac{1}{\pi}\frac{m_\sigma \Gamma_\sigma}{(s-m_\sigma^2)^2 + m_\sigma^2 \Gamma_\sigma^2}\,
 \frac{p^*(\sqrt{s},m_\pi,m_\pi)}{p^*(m_\sigma,m_\pi,m_\pi)}\,\frac{s}{m^2_\sigma}\,\BR\left( \sigma \to \pi^+ \pi^-\right),
\end{align}\label{eq:widths}\end{subequations}where $p^*(m_1,m_2,m_3)$ is the decay 3-momentum in the $m_1$ rest frame. In Eq.~\eqref{widthZ}, the factor of  $2$ takes into account the incoherent sum over the two charged resonances, being the interference numerically negligible. 

For the sake of simplicity, since we are not able to resolve the details of the lineshape within our large uncertainties, we considered the $\sigma$ resonance to be described by a Breit-Wigner distribution with mass and width \mbox{$M_\sigma = (475 \pm 75)$~MeV}, $\Gamma_\sigma = (550 \pm 150)$~MeV.
To obtain the branching ratio $ \BR\left(Z^\prime_c \to h_c \,\pi\right)$, we assume the total width of $Z_c^\prime$ to be saturated by the observed decay modes into $h_c \pi$~\cite{beshcpipi} and $D^* \bar D^*$~\cite{besDstarDstar}. We use the BES measurements of production cross sections
\begin{align}
\sigma(e^+e^- \to Z_c^{\prime\pm}\pi^{\mp} \to h_c \pi^{+} \pi^{-}) &=\left(7.4\pm1.7\pm2.1\pm1.2\right)\,\text{pb},\\
\sigma(e^+e^- \to (D^*\bar D^*)^{\pm}\pi^{\mp})&=\left( 137\pm9\pm15 \right) \,\text{pb},
\end{align}
and of the cross sections ratio
\begin{equation}
R=\frac{\sigma(e^+e^- \to Z_c^{\prime\pm} \pi^{\mp} \to (D^*\bar D^*)^{\pm}\pi^{\mp})}{\sigma(e^+e^- \to (D^*\bar D^*)^{\pm}\pi^{\mp})}=0.65\pm0.09\pm0.06,
\end{equation}
to estimate the branching ratio
\begin{align}
\BR(Z_c^\prime\to h_c\pi) &=(8.0\pm3.6) \%.
\end{align}
The branching fraction $\BR(\sigma\to\pi^+\pi^-)$ can be assumed to be $\simeq \frac{2}{3}$ via isospin symmetry. 
The effective strong couplings $g_\chi$, $g_h$, $g_Z$ in Eq.~\eqref{eq:widths} are unknown and should be fitted from data. 

To obtain a prediction within the diquark-antidiquark model, we assume that a tetraquark couples universally to any charmonia, {\it i.e.} that the strong effective couplings are equal to a universal constant times a factor depending on heavy quark spin content~\cite{Maiani:2014aja,alimaiani,etacrho}.

In the $\ket{s_{c\bar c} ,  s_{q\bar q}}$ basis, we have:
\begin{align}
\ket{Y(4220)}&=\frac{\sqrt{3}}{2}\ket{0,0}-\frac{1}{2}\ket{1,1},\nonumber\\
\ket{Z_c^\prime}&=\frac{1}{\sqrt{2}}\left(\ket{1,0}+\ket{0,1}\right) 
\end{align}
and we recall
\begin{align}
\ket{h_c} &= \ket{s_{c\bar c} = 0}, & \ket{\chi_{cJ}} &= \ket{s_{c\bar c} = 1}.
\end{align}
Hence, we get $g_h : g_\chi = \braket{Y|h_c} : \braket{Y | \chi_{cJ}} = \sqrt{3} : 1$.
The estimate of the ratio $g_Z : g_\chi$ deserves a separate comment. The decay $Y(4220) \to Z_c^\prime \pi$ is an hadronic transition between tetraquark states. 
With the additional assumption that the dynamics of tetraquark transitions is the same as that of tetraquark-charmonium decays, one could get 
\mbox{$g_Z : g_\chi = \left\langle  Y | Z_c^\prime \right\rangle : \braket{Y|\chi_{cJ}}  = \frac{\sqrt{3}-1}{2\sqrt{2}}:\frac{1}{2} \simeq 0.52$.} 
This result is potentially affected by large corrections. 
Comparisons with new tetraquark candidates decays will allow us to probe the validity of this assumption, and evaluate the errors properly.  
That said, an order-of-magnitude estimate is given by the ratio:  

\begin{figure}[t]
\centering
\includegraphics[width=0.55\textwidth]{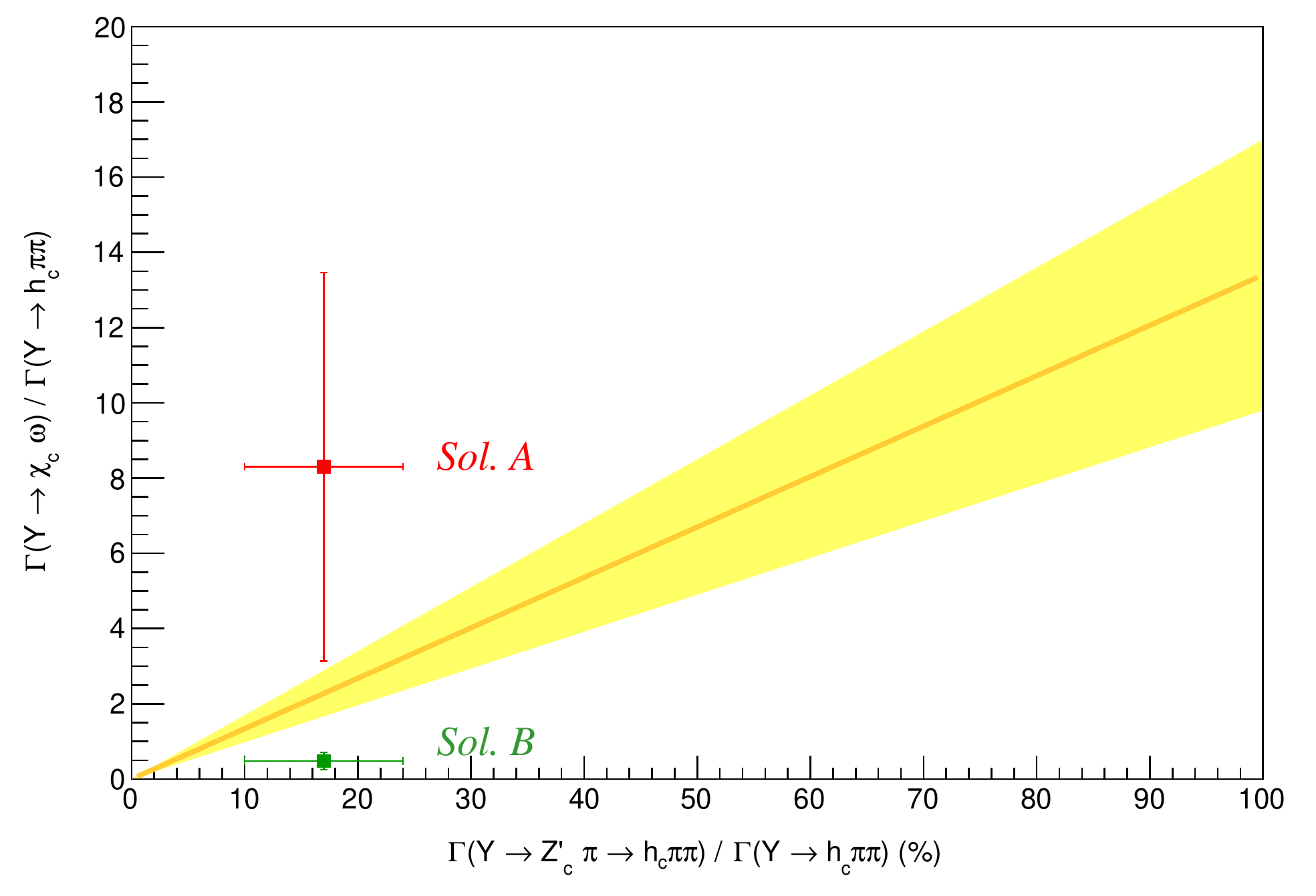}
\caption{Measurements and predictions for $\Gamma\left(Y\to \chicomega\right)/\Gamma\left(Y\to\hcpipi\right)$ as a function of $R_{YZ} = \sigma\left(Y \to Z_c^{\prime\pm} \pi^\mp \to h_c \pi^+ \pi^-\right) / \sigma\left(Y \to \hcpipi\right)$. The solid line is the prediction as a function of $R_{YZ}$, the colored band the correspondent error. The red and green points are the experimental ratios in Eq.~\eqref{ratexp}, plotted at the measured $R_{YZ} = (17 \pm 7)\%$. }
\label{fig:perc}
\end{figure}

\begin{subequations}
\begin{equation}
\frac{\Gamma\left(Y(4220)\to\chicomega\right)}{\Gamma(Y(4220)\to Z_c^{\prime\pm}\pi^{\mp}\to \hcpipi)} = 13.4 \pm 3.6,\label{eq:100}
\end{equation}
hence
\begin{equation}
\frac{\Gamma\left(Y(4220)\to\chicomega\right)}{\Gamma(Y(4220)\to \hcpipi)} = \left(13.4 \pm 3.6\right)\times R_{YZ} = 2.3 \pm 1.2.
\end{equation}%
In \figurename{~\ref{fig:perc}} we show this result as a function of $R_{YZ}$. For the quoted value of $R_{YZ}$, the ratio is compatible with the solution~\eqref{ratexpbuono} of the fit, even better if $R_{YZ}$ will be discovered to be larger. 
Similarly, we predict
\begin{equation}
 \frac{\Gamma\left(Y(4220)\to Z_c^{\prime\pm}\pi^\mp \to \hcpipi\right)}{\Gamma\left(Y(4220)\to h_c \sigma \to \hcpipi\right)} = 4.8 \pm 3.5,
\end{equation}%
\label{eq:Bsigma}\end{subequations} which can be verified by a detailed Dalitz analysis when more data will be available.
The errors in Eq.~\eqref{eq:Bsigma} are due to the experimental uncertainty on masses, widths and branching fractions of the intermediate resonances. 
We stress that we are not considering the error on the couplings.

In conclusion, the structures seen by BES~III in \hcpipi and \chicomega can be explained within the diquark-antidiquark model.  
The results are summarized in \figurename{~\ref{fig:perc}}. A detailed analysis of the $e^+ e^- \to \hcpipi$ Dalitz plot as a function of $\sqrt{s}$ will establish the value of $R_{YZ}$ and further constrain this model when more data from BES~III will be available. 

\begin{acknowledgments}
We wish to thank L.~Maiani and V.~Riquer for many interesting comments and discussions. We also thank C.P.~Shen for useful comments on the first version of this paper.
\end{acknowledgments}


\begin{thebibliography}{10}
\bibitem{beschicomega}
  M.~Ablikim {\it et al.}  [BESIII Collaboration],
  Phys.\ Rev.\ Lett.\  {\bf 114} (2015) 9,  092003
  [\href{http://arxiv.org/abs/1410.6538}{arXiv:1410.6538 [hep-ex]}].  
\bibitem{altri}
X.~Li and M.~B.~Voloshin,
Phys.\ Rev.\ D {\bf 91}, 034004 (2015)
[\href{http://arxiv.org/abs/1411.2952}{arXiv:1411.2952 [hep-ph]}];
D.~Y.~Chen, X.~Liu and T.~Matsuki,
\href{http://arxiv.org/abs/1411.5136}{arXiv:1411.5136 [hep-ph]}.
  
\bibitem{beshcpipi}
  M.~Ablikim {\it et al.}  [BESIII Collaboration],
  Phys.\ Rev.\ Lett.\  {\bf 111} (2013) 24,  242001
  \href{http://arxiv.org/abs/1309.1896}{arXiv:1309.1896 [hep-ex]}.
\bibitem{hadrocharmonium} 
  X.~Li and M.~B.~Voloshin,
  Mod.\ Phys.\ Lett.\ A {\bf 29}, no. 12, 1450060 (2014)
  [\href{http://arxiv.org/abs/1309.1681}{arXiv:1309.1681 [hep-ph]}].

  \bibitem{yuan}
  C.~Z.~Yuan,
  Chin.\ Phys.\ C {\bf 38} (2014) 043001
  [\href{http://arxiv.org/abs/1312.6399}{arXiv:1312.6399 [hep-ex]}]; 
  Int.\ J.\ Mod.\ Phys.\ A {\bf 29} (2014) 1430046
  [\href{http://arxiv.org/abs/1404.7768}{arXiv:1404.7768 [hep-ex]}].

  \bibitem{HQSvoloshin}
  A.~E.~Bondar, A.~Garmash, A.~I.~Milstein, R.~Mizuk and M.~B.~Voloshin,
  Phys.\ Rev.\ D {\bf 84}, 054010 (2011)
  [\href{http://arxiv.org/abs/1105.4473}{arXiv:1105.4473 [hep-ph]}].
  
  \bibitem{alivecchio}
  A.~Ali, C.~Hambrock and W.~Wang,
  Phys.\ Rev.\ D {\bf 85}, 054011 (2012)
  [\href{http://arxiv.org/abs/1110.1333}{arXiv:1110.1333 [hep-ph]}].

  \bibitem{alimaiani}
  A.~Ali, L.~Maiani, A.~D.~Polosa and V.~Riquer,
  Phys.\ Rev.\ D {\bf 91}, 017502 (2015)
  [\href{http://arxiv.org/abs/1412.2049}{arXiv:1412.2049 [hep-ph]}].

\bibitem{tetraquarks} 
L.~Maiani, F.~Piccinini, A.~D.~Polosa and V.~Riquer,
  Phys.\ Rev.\ D {\bf 71}, 014028 (2005)
  [\href{http://arxiv.org/abs/hep-ph/0412098}{hep-ph/0412098}];
  L.~Maiani, V.~Riquer, R.~Faccini, F.~Piccinini, A.~Pilloni and A.~D.~Polosa,
  Phys.\ Rev.\ D {\bf 87}, no. 11, 111102 (2013)
  [\href{http://arxiv.org/abs/1303.6857}{arXiv:1303.6857 [hep-ph]}].

\bibitem{review} 
  A.~Esposito, A.~L.~Guerrieri, F.~Piccinini, A.~Pilloni and A.~D.~Polosa,
  Int.\ J.\ Mod.\ Phys.\ A {\bf 30}, 1530002 (2015)
  [\href{http://arxiv.org/abs/1411.5997}{arXiv:1411.5997 [hep-ph]}];
    R.~Faccini, A.~Pilloni and A.~D.~Polosa,
  Mod.\ Phys.\ Lett.\ A {\bf 27}, 1230025 (2012)
  [\href{http://arxiv.org/abs/1209.0107}{arXiv:1209.0107 [hep-ph]}];
    N.~Drenska, R.~Faccini, F.~Piccinini, A.~Polosa, F.~Renga and C.~Sabelli,
  Riv.\ Nuovo Cim.\  {\bf 33}, 633 (2010)
  [\href{http://arxiv.org/abs/1006.2741}{arXiv:1006.2741 [hep-ph]}].
  \bibitem{Maiani:2014aja}
      L.~Maiani, F.~Piccinini, A.D.~Polosa, V.~Riquer,
      Phys.\ Rev.\  {\bf D89} (2014) 114010
      [\href{http://arxiv.org/abs/1405.1551}{arXiv:1405.1551 [hep-ph]}].
    
\bibitem{cleohcpipi}
  T.~K.~Pedlar {\it et al.}  [CLEO Collaboration],
  Phys.\ Rev.\ Lett.\  {\bf 107} (2011) 041803
  [\href{http://arxiv.org/abs/1104.2025}{arXiv:1104.2025 [hep-ex]}].
  
\bibitem{soviet} 
  M.~B.~Voloshin,
  Sov.\ J.\ Nucl.\ Phys.\  {\bf 43}, 1011 (1986)
  [Yad.\ Fiz.\  {\bf 43}, 1571 (1986)].

      \bibitem{PC}
  	C.~Z.~Yuan,
	private communication.
  
\bibitem{feshbach} 
  M.~Papinutto, F.~Piccinini, A.~Pilloni, A.~D.~Polosa and N.~Tantalo,
  \href{http://arxiv/abs/1311.7374}{arXiv:1311.7374 [hep-ph]};
  A.~L.~Guerrieri, F.~Piccinini, A.~Pilloni and A.~D.~Polosa,
  Phys.\ Rev.\ D {\bf 90}, 034003 (2014)
  [\href{http://arxiv/abs/1405.7929}{arXiv:1405.7929 [hep-ph]}].
%
\bibitem{strong}
  F.~Brazzi, B.~Grinstein, F.~Piccinini, A.~D.~Polosa and C.~Sabelli,
  Phys.\ Rev.\ D {\bf 84}, 014003 (2011)
  [\href{http://arxiv.org/abs/1103.3155}{arXiv:1103.3155 [hep-ph]}].
 
\bibitem{besDstarDstar} 
  M.~Ablikim {\it et al.}  [BESIII Collaboration],
  Phys.\ Rev.\ Lett.\  {\bf 112}, no. 13, 132001 (2014)
  \href{http://arxiv.org/abs/1308.2760}{arXiv:1308.2760 [hep-ex]}.

\bibitem{etacrho} 
  A.~Esposito, A.~L.~Guerrieri and A.~Pilloni,
  Phys.\ Lett.\ B {\bf 746}, 194-201 (2015)
  [\href{http://arxiv.org/abs/1409.3551}{arXiv:1409.3551 [hep-ph]}].
\end{thebibliography}
\end{document}